\documentclass{article}
\usepackage{multicol}
\usepackage{graphicx}
\usepackage{amsmath}
\usepackage{amssymb}
\usepackage{url}
\usepackage{units}
\usepackage{comment}
\usepackage{xcolor}
\usepackage{siunitx}
\usepackage{subcaption}
\usepackage{lineno}
\usepackage[utf8]{inputenc}
\usepackage[T1,T2A]{fontenc}
\usepackage[backend=bibtex, defernumbers=true, style=nature]{biblatex}
\usepackage{datetime2}

\usepackage[a4paper, scale=0.84]{geometry}
\usepackage{newfloat}

\usepackage{caption}
\DeclareFloatingEnvironment[
    fileext=loed,
    listname={List of Extended Data},
    name=Extended Data Figure,
    placement=tbhp,
    within=section,
]{edfigure}
\DeclareFloatingEnvironment[
    fileext=loet,
    listname={List of Extended Table},
    name=Extended Data Table,
    placement=tbhp,
    within=section,
]{etable}

\makeatletter
\AtEveryCitekey{%
  \ifcsundef{blx@entry@refsegment@\the\c@refsection @\thefield{entrykey}}
    {\csnumgdef{blx@entry@refsegment@\the\c@refsection @\thefield{entrykey}}{\the\c@refsegment}}
    {}}
\defbibcheck{onlynew}{%
  \ifnumless{0\csuse{blx@entry@refsegment@\the\c@refsection @\thefield{entrykey}}}{\the\c@refsegment}
    {\skipentry}
    {}}
\makeatother

\DeclareFieldFormat*{title}{#1}
\DeclareFieldFormat[inproceedings]{title}{#1}

\DTMsavetimestamp{special}{2016-12-08T01:47:59+00:00}
\newcommand{\nuebar}{\bar{\nu}_e}
\newcommand{\depe}{$6.05 \pm 0.72$\,PeV}

\newcommand{\mue}{$26.4_{-12.4}^{+28.6}$\,GeV}
\newcommand{\like}{\mathcal{L}}
\newcommand{\prob}{\mathcal{P}}
\usepackage{fancyhdr}
\begin{document}
\title{Detection of a particle shower at the Glashow resonance with IceCube\footnote{\textit{Nature} 591, 220–224 (2021). \protect\url{https://doi.org/10.1038/s41586-021-03256-1}}
}
\date{\vspace{-5ex}}
\author{}
\maketitle
\begin{minipage}{0.95\textwidth}
\small
\author{IceCube Collaboration: M. G. Aartsen$^{17}$, R. Abbasi$^{16}$, M. Ackermann$^{56}$, J. Adams$^{17}$, J. A. Aguilar$^{12}$, M. Ahlers$^{21}$, M. Ahrens$^{47}$, C. Alispach$^{27}$, N. M. Amin$^{40}$, K. Andeen$^{38}$
% ...
}
\end{minipage}
\newpage
{
\small
\begin{enumerate}
\item III. Physikalisches Institut, RWTH Aachen University, D-52056 Aachen, Germany
\item Department of Physics, University of Adelaide, Adelaide, 5005, Australia
\item Dept. of Physics and Astronomy, University of Alaska Anchorage, 3211 Providence Dr., Anchorage, AK 99508, USA
\item Dept. of Physics, University of Texas at Arlington, 502 Yates St., Science Hall Rm 108, Box 19059, Arlington, TX 76019, USA
\item CTSPS, Clark-Atlanta University, Atlanta, GA 30314, USA
\item School of Physics and Center for Relativistic Astrophysics, Georgia Institute of Technology, Atlanta, GA 30332, USA
\item Dept. of Physics, Southern University, Baton Rouge, LA 70813, USA
\item Dept. of Physics, University of California, Berkeley, CA 94720, USA
\item Lawrence Berkeley National Laboratory, Berkeley, CA 94720, USA
\item Institut für Physik, Humboldt-Universität zu Berlin, D-12489 Berlin, Germany
\item Fakultät für Physik \& Astronomie, Ruhr-Universität Bochum, D-44780 Bochum, Germany
\item Université Libre de Bruxelles, Science Faculty CP230, B-1050 Brussels, Belgium
\item Vrije Universiteit Brussel (VUB), Dienst ELEM, B-1050 Brussels, Belgium
\item Dept. of Physics, Massachusetts Institute of Technology, Cambridge, MA 02139, USA
\item Dept. of Physics and Institute for Global Prominent Research, Chiba University, Chiba 263-8522, Japan
\item Department of Physics, Loyola University Chicago, Chicago, IL 60660, USA
\item Dept. of Physics and Astronomy, University of Canterbury, Private Bag 4800, Christchurch, New Zealand
\item Dept. of Physics, University of Maryland, College Park, MD 20742, USA
\item Dept. of Astronomy, Ohio State University, Columbus, OH 43210, USA
\item Dept. of Physics and Center for Cosmology and Astro-Particle Physics, Ohio State University, Columbus, OH 43210, USA
\item Niels Bohr Institute, University of Copenhagen, DK-2100 Copenhagen, Denmark
\item Dept. of Physics, TU Dortmund University, D-44221 Dortmund, Germany
\item Dept. of Physics and Astronomy, Michigan State University, East Lansing, MI 48824, USA
\item Dept. of Physics, University of Alberta, Edmonton, Alberta, Canada T6G 2E1
\item Erlangen Centre for Astroparticle Physics, Friedrich-Alexander-Universität Erlangen-Nürnberg, D-91058 Erlangen, Germany
\item Physik-department, Technische Universität München, D-85748 Garching, Germany
\item Département de physique nucléaire et corpusculaire, Université de Genève, CH-1211 Genève, Switzerland
\item Dept. of Physics and Astronomy, University of Gent, B-9000 Gent, Belgium
\item Dept. of Physics and Astronomy, University of California, Irvine, CA 92697, USA
\item Karlsruhe Institute of Technology, Institut für Kernphysik, D-76021 Karlsruhe, Germany
\item Dept. of Physics and Astronomy, University of Kansas, Lawrence, KS 66045, USA
\item SNOLAB, 1039 Regional Road 24, Creighton Mine 9, Lively, ON, Canada P3Y 1N2
\item Department of Physics and Astronomy, UCLA, Los Angeles, CA 90095, USA
\item Department of Physics, Mercer University, Macon, GA 31207-0001, USA
\item Dept. of Astronomy, University of Wisconsin–Madison, Madison, WI 53706, USA
\item Dept. of Physics and Wisconsin IceCube Particle Astrophysics Center, University of Wisconsin–Madison, Madison, WI 53706, USA
\item Institute of Physics, University of Mainz, Staudinger Weg 7, D-55099 Mainz, Germany
\item Department of Physics, Marquette University, Milwaukee, WI, 53201, USA
\item Institut für Kernphysik, Westfälische Wilhelms-Universität Münster, D-48149 Münster, Germany
\item Bartol Research Institute and Dept. of Physics and Astronomy, University of Delaware, Newark, DE 19716, USA
\item Dept. of Physics, Yale University, New Haven, CT 06520, USA
\item Dept. of Physics, University of Oxford, Parks Road, Oxford OX1 3PU, UK
\item Dept. of Physics, Drexel University, 3141 Chestnut Street, Philadelphia, PA 19104, USA
\item Physics Department, South Dakota School of Mines and Technology, Rapid City, SD 57701, USA
\item Dept. of Physics, University of Wisconsin, River Falls, WI 54022, USA
\item Dept. of Physics and Astronomy, University of Rochester, Rochester, NY 14627, USA
\item Oskar Klein Centre and Dept. of Physics, Stockholm University, SE-10691 Stockholm, Sweden
\item Dept. of Physics and Astronomy, Stony Brook University, Stony Brook, NY 11794-3800, USA
\item Dept. of Physics, Sungkyunkwan University, Suwon 16419, Korea
\item Institute of Basic Science, Sungkyunkwan University, Suwon 16419, Korea
\item Dept. of Physics and Astronomy, University of Alabama, Tuscaloosa, AL 35487, USA
\item Dept. of Astronomy and Astrophysics, Pennsylvania State University, University Park, PA 16802, USA
\item Dept. of Physics, Pennsylvania State University, University Park, PA 16802, USA
\item Dept. of Physics and Astronomy, Uppsala University, Box 516, S-75120 Uppsala, Sweden
\item Dept. of Physics, University of Wuppertal, D-42119 Wuppertal, Germany
\item DESY, D-15738 Zeuthen, Germany
\item[a ] also at Università di Padova, I-35131 Padova, Italy
\item[b ] also at National Research Nuclear University, Moscow Engineering Physics Institute (MEPhI), Moscow 115409, Russia
\item[c ] also at Earthquake Research Institute, University of Tokyo, Bunkyo, Tokyo 113-0032, Japan

\end{enumerate}
}

\bigskip
\bigskip

\newrefsegment 

\begin{abstract}
The Glashow resonance describes the resonant formation of a $W^-$ boson during the interaction of a high-energy electron antineutrino with an electron~\cite{Glashow:1960zz}, peaking at an antineutrino energy of 6.3 petaelectronvolts (PeV) in the rest frame of the electron. Whereas this energy scale is out of reach for currently operating and future planned particle accelerators, natural astrophysical phenomena are expected to produce antineutrinos with energies beyond the PeV scale. Here we report the detection by the IceCube neutrino observatory of a cascade of high-energy particles (a particle shower) consistent with being created at the Glashow resonance. A shower with an energy of \depe{} (determined from Cherenkov radiation in the Antarctic Ice Sheet) was measured. Features consistent with the production of secondary muons in the particle shower indicate the hadronic decay of a resonant $W^-$ boson, confirm that the source is astrophysical and provide improved directional localization. The evidence of the Glashow resonance suggests the presence of electron antineutrinos in the astrophysical flux, while also providing further validation of the standard model of particle physics. Its unique signature indicates a method of distinguishing neutrinos from antineutrinos, thus providing a way to identify astronomical accelerators that produce neutrinos via hadronuclear or photohadronic interactions, with or without strong magnetic fields. As such, knowledge of both the flavour (that is, electron, muon or tau neutrinos) and charge (neutrino or antineutrino) will facilitate the advancement of neutrino astronomy.
\end{abstract}

\bigskip
\bigskip

In this Article we present a search for very-high-energy astrophysical neutrinos with IceCube. One event was found with a visible energy of \depe{}. Given its energy and direction, it is classified as an astrophysical neutrino at the $5\sigma$ level. Furthermore, data collected by the sensors closest to the interaction point, as well as the measured energy, are consistent with the hadronic decay of a $W^-$ boson produced on the Glashow resonance. Taking into account only the detector’s energy resolution, the probability that the event is produced off-resonance by deep inelastic scattering is 0.01 assuming the best-fit flux from~\cite{Mohrmann:2015mgs}.
The neutrino energy is inferred to be about 6.3 PeV by correcting the visible energy for shower particles that do not radiate.

Neutrinos are fundamental particles that couple to matter only via $W^{\pm}$ or $Z^0$ boson exchange. As such, they are uniquely suitable messengers to study high-energy particle accelerators in the universe because they can escape dense media surrounding the production region without interaction and travel to Earth without being deflected by magnetic fields. In the interaction of electron antineutrinos ($\nuebar$) with electrons, the standard model predicts the s-channel production of a $W^-$ boson. For a centre-of-mass energy $\sqrt{s}=M_W=\SI{80.38}{\giga \eV}$ (the mass of the $W^-$) the cross section becomes resonantly enhanced~\cite{Glashow:1960zz}. The standard model cross-section, $\sigma(s)$, for the process $\nuebar+e^- \rightarrow W^- \rightarrow X$ is: 
\begin{equation}
    \sigma(s) = 24\pi \Gamma_W^2 \cdot B_{W^- \rightarrow \nuebar+e^-} \frac{s / M_W^2}{(s-M_W^2)^2+\Gamma_W^2 M_W^2}\, ,
    \label{eq:gr}
\end{equation}
where $\Gamma_W=\SI{2.09}{\giga \eV}$ is the $W^-$ decay width and $B_{W^- \rightarrow \nuebar+e^-}$ its branching ratio for the indicated channel~\cite{Barger:2014iua,PDG}. It is clear from Eq.~(\ref{eq:gr}) that $\sigma(s)$ is maximal when $s=M^{2}_W$.
In the electron (mass $m_e = \SI{0.511}{\MeV}$) rest frame, the resonance energy is $E_R = M_W^2/(2m_{e}) = \SI{6.32}{\peta \eV}$.

The resonance energy lies beyond terrestrial accelerators, but not astrophysical sources of neutrinos. Additionally, since the Glashow resonance occurs only for $\nuebar$, it is a unique probe of the production mechanism. Neutrinos are expected to be produced in the interaction of high-energy cosmic rays (typically protons) with matter or ambient radiation. In the simplest proton-photon (p$\gamma$) interaction source model, without multi-pion production, the ratio $\bar{\nu}_{e}:\nu_{e}=1:3.5$ at Earth~\cite{Barger:2014iua}. 
If, however, there is also a strong magnetic field,
$B \geq \SI{0.033}{\tesla}\times \eta/(1+z)$, where $z$ is redshift and $\eta$ is Lorentz boost of the source, synchrotron losses start to dominate over muon decay. This prevents the creation of $\bar{\nu}_{e}$, which results in a near-zero $\bar{\nu}_{e}:\nu_{e}$ ratio at Earth~\cite{Barger:2014iua,Kashti:2005qa}. In the proton-proton (pp) interaction source model, in which cosmic rays interact with the background gas to generate an approximately equal mixture of $\pi^0$, $\pi^-$ and $\pi^+$, one expects $\bar{\nu}_{e}:\nu_{e}=1:1$ at Earth. A statistically significant measurement of the Glashow resonance event rate directly probes the antineutrino fraction and thus helps to constrain neutrino production mechanism(s).

\begin{figure}[htp]
\centering
\includegraphics[width=.75\textwidth]{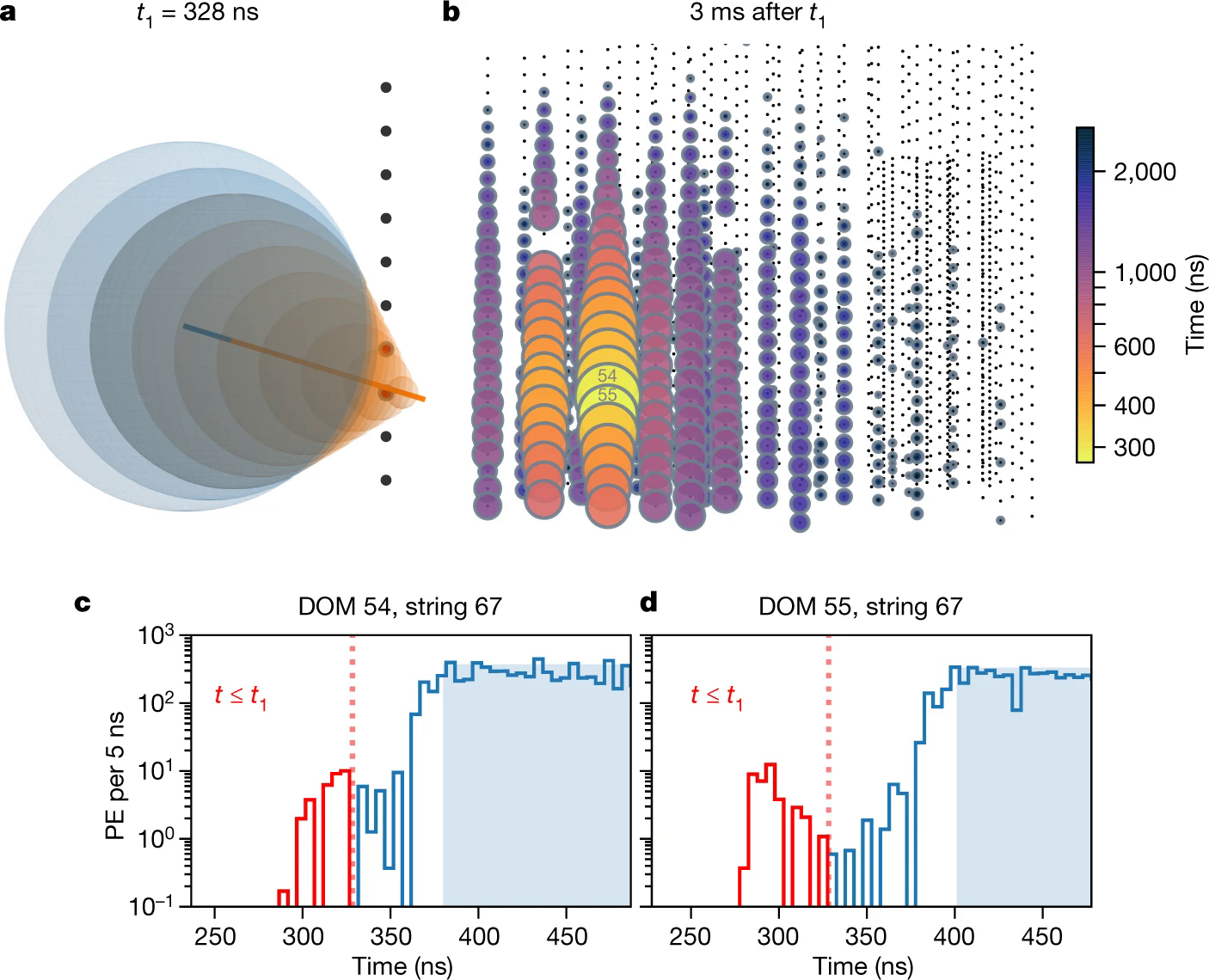}
\caption{\textbf{Visualization of detected photons at different times and distribution of early pulses.} \textbf{a}, Schematic of an escaping muon travelling at faster than the speed of light (in ice) and its Cherenkov cone (orange). The muons reach the nearest modules (DOMs 54 and 55 on string 67) ahead of the Cherenkov photons produced by the EM component of the hadronic shower (blue) as these travel at the speed of light in ice. The blue line is associated with the average distance travelled by the main shower, while the orange line extends further and is associated with the muons. Each black dot arranged vertically is a DOM on the nearest string, with the two (slightly larger) dots inside the orange cone the first two to observe early pulses. The time $t_{1}$ indicates the approximate time elapsed since the neutrino interaction at which this snapshot graphic was taken. \textbf{b}, Event view, showing DOMs that triggered across IceCube at a later time. Each bubble represents a DOM, with its size proportional to the deposited charge. Colours indicate the time each DOM first triggered, relative to our best knowledge of when the initial interaction occurred. The small black dots are DOMs further away that did not detect photons \SI{3}{\ms} after $t_{1}$.
\textbf{c, d,} Distributions of the deposited charge over time on the two earliest hit DOMs, 54 (\textbf{c}) and 55 (\textbf{d}). The dotted red line is at $t_{1} = \SI{328}{ns}$, the instant shown in \textbf{a}. The histogram in red (blue) shows photons arriving before (after) $t_{1}$, and the blue shaded region denotes saturation of the photomultiplier tube. \label{fig:event_view}}
\end{figure}

As the flux of astrophysical neutrinos falls off following a power law in energy~\cite{Bell:1978zc} and its intensity is bounded by cosmic-ray observations~\cite{Waxman:1998yy}, a large-volume detector is needed to detect PeV neutrinos. The IceCube neutrino observatory, situated at the geographic South Pole, instruments a cubic kilometre of ice \SIrange{1450}{2450}{\m} beneath the surface~\cite{Aartsen:2016nxy}—a natural detection medium. It has measured the flux of neutrinos between $\SI{10}{GeV}$ and $\SI{10}{PeV}$, and is sensitive to neutrinos beyond $\SI{1}{EeV}$. As neutrinos are uncharged, they are detected in IceCube by the Cherenkov radiation from secondary charged particles produced by their interactions. Cherenkov light collected by digital optical modules (DOMs) is used to reconstruct properties such as the visible energy and incoming direction of the primary neutrino~\cite{Chirkin:2013,Aartsen:2013vja}. 
The visible energy is defined as the energy required of an electromagnetic (EM) shower to produce the light yield observed. As it has no magnet, IceCube cannot distinguish between neutrino and antineutrino interactions on the basis of the charge of the outgoing lepton—whether neutrinos are Dirac or Majorana particles (the latter implying that they are their own antiparticles) remains unresolved. However, owing to the good timing resolution~\cite{IceCube:2008qbc,Abbasi:2010vc} (about $\SI{2}{ns}$) of the DOMs , the structure of waveforms recorded by individual modules may contain additional information on the event~\cite{Aartsen:2015dlt}.

A machine-learning-based algorithm was run to obtain a sample of PeV energy partially contained events (PEPEs)~\cite{Lu:2017nti}. By selecting events near the edge of the detector, the detection volume is increased compared to previous analyses that rely on a smaller, central fiducial volume. Data from May 2012 to May 2017, corresponding to a total live-time of 4.6 years, were analysed. 
One event was detected on 2016 December 8 at 01:47:59 UTC with visible energy greater than $\SI{4}{PeV}$, which is an energy threshold well below the resonance energy and chosen \emph{a posteriori} in order to study this particular event. The event is shown in Fig.\,\ref{fig:event_view}, with a reconstructed vertex approximately $\SI{80}{m}$ from the nearest DOM. 
The same event was also found in the 9-year extremely high energy search~\cite{Aartsen:2018vtx}. Accounting for systematic uncertainties in photon propagation due to the ice model—a parameterization of the scattering and absorption lengths of light in the ice~\cite{Aartsen:2013rt}—and the overall detector calibration, the visible energy of the event is \depe{}. This is consistent with a $\SI{6.3}{PeV}$ $W^-$ that decays hadronically, since roughly 5\% of that energy is expected to be taken by particles that do not emit detectable Cherenkov radiation~\cite{Aartsen:2013vja}. The boosted decision tree (BDT) classification score is well above the signal threshold, and \emph{a posteriori} studies of this event, discussed below, lead us to conclude that the event is very likely to be of astrophysical origin.

The main shower was reconstructed by repeating Monte Carlo (MC) simulations under different parameters to find the best-fit energy, vertex and direction~\cite{Chirkin:2013}.
By varying the ice model used in the reconstruction, detector systematic uncertainties on the visible energy, direction and vertex position of the shower were evaluated. Additionally, a global energy scale uncertainty associated with the overall detector calibration was applied to the energy reconstruction.

After reconstruction, the three DOMs closest to the reconstructed vertex were found to have detected pulses earlier than is possible for photon traveling in ice in ice at $v=\SI{2.19e8}{\m \per \s}$. 
Such pulses can, however, be produced by muons created from meson decays in the hadronic shower, which travel close to the speed of light in vacuum ($c=\SI{3.00e8}{\m \per \s}$). These muons outrun the Cherenkov wavefront of the main shower (by about \SI{1.23}{\nano \s \per \m}) while producing Cherenkov radiation near the DOMs, thus depositing early pulses in them, as illustrated in Fig.\,\ref{fig:event_view}a.

\begin{figure}
\centering
\includegraphics[width=0.8\textwidth]{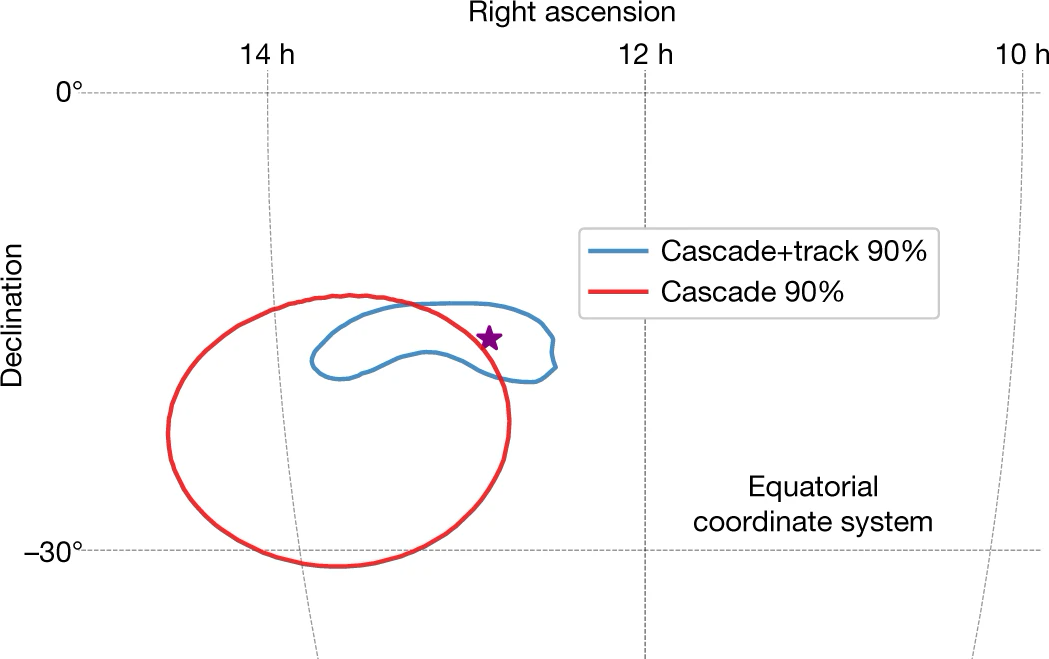}
\caption{\textbf{Directional reconstructions under two hypotheses.} The 90\% contours from the cascade (red) and hybrid cascade+track (blue) directional reconstructions are shown in equatorial coordinates. The most-probable direction according to the hybrid reconstruction is shown as the purple star. Systematic uncertainties on the scattering and absorption of photons in ice have been included in both contours. The results are consistent and indicate a common origin. The hybrid reconstruction improves pointing, reducing the contour area by a factor of 5, and should be beneficial to the multi-messenger campaign for such events in the future. Effects of ice anisotropy are shown in the Methods section, and the combined best-fit direction is at right ascension (RA) 12 h 50 min 47.9 s and declination (dec.) $-$15.9$^{\circ}$. The area within the 90\% probability cascade+track contour is about 68$\mathrm{deg}^2$ probability cascade+track contour is about 68$\mathrm{deg}^2$.}
\label{fig:rec_direction}
\end{figure}

A second reconstruction using only the early pulses to fit a track hypothesis further improves and verifies the directional reconstruction of this event.
The two reconstructed directions agree within uncertainties, as shown in Fig.\,\ref{fig:rec_direction}. This indicates that the muons and hadronic shower travel along the same general direction, as is expected from relativistic kinematics. On the Basis of the observation that early pulses occurred only on the nearest string, a most-probable leading muon energy of \mue{} was obtained. This is consistent with a distribution of leading muon energies from MC simulations of a \SI{6.3}{\peta \eV} hadronic shower, which has quartiles of $(20, 37, 72)$\,GeV. 

Information from both reconstructions refines the estimate of expected backgrounds compared to the sample average. 
The only possibility for a cosmic-ray-induced atmospheric muon to produce both a \SI{6}{PeV} cascade and early pulses, as in this event, is for it to reach IceCube at PeV energies and deposit nearly all its energy over a few metres. As a conservative estimate, this background rate was numerically calculated by considering all atmospheric muons that intersect a cylinder centred on IceCube with radius 800\,m and height 1600\,m. 
By then requiring that muons deposit a visible energy similar to that of the cascade over a short distance, but retain the energy allowed by early pulses, the background muon flux is further reduced to give an expectation rate of $1.1\times 10^{-7}$ events in 4.6 years. This allows an \emph{a posteriori} rejection of the cosmic-ray muon background hypothesis by over $5\sigma$.

Similarly, the early pulse signature can be used to reject the atmospheric neutrino background hypothesis. 
Above roughly \SI{100}{\tera \eV}, the atmospheric neutrino flux from the prompt decay of charmed mesons is expected to be greater than that from the decay of longer-lived pions and kaons. 
Charmed mesons can decay to electron (anti)neutrinos that are often accompanied by muons produced in other branches of the same air shower. These muons can be used to veto atmospheric neutrinosr~\cite{Gaisser:2014bja,Arguelles:2018awr}.
The expectation rate of atmospheric neutrinos passing the PEPE event selection with accompanying muon energy consistent with the observed early pulses is around $2\times 10^{-7}$ in 4.6 years. We conclude that the event is induced by an astrophysical neutrino.

\begin{figure}[htp]
\centering
\includegraphics[width=.6 \textwidth]{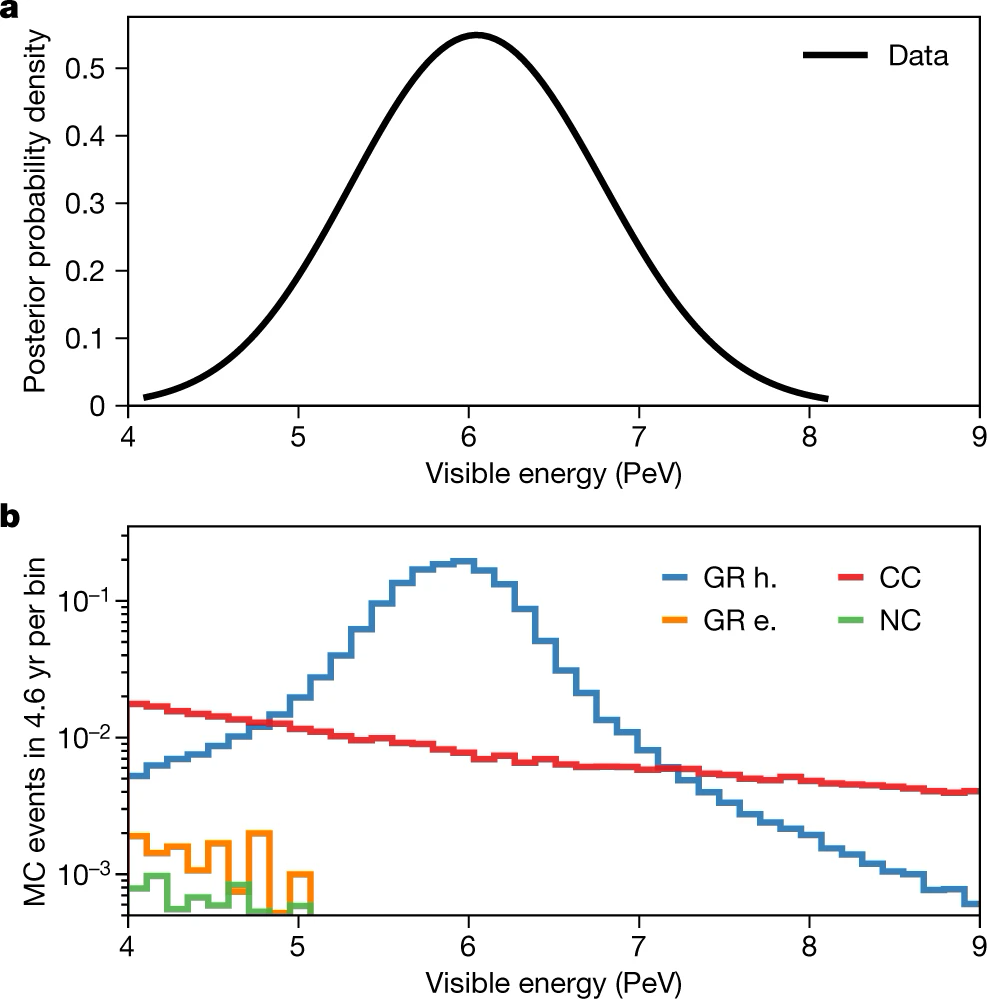}

    \caption{\textbf{Reconstructed energy posterior probability density and expected distributions from MC simulations.} \textbf{a}, 
    Posterior probability density of the visible energy for this event. Systematic uncertainties due to the ice and global energy scale of the detector are included. 
    \textbf{b}, Expected Monte Carlo (MC) event distributions in visible energy of hadrons from $W^-$ decay (GR h., blue), the electron from $W^-$ decay (GR e., orange), charged-current interactions (CC; red) and neutral-current interactions (NC; green) for a live-time of 4.6 years from the PEPE sample. We assume the ratio $\bar{\nu}:\nu=1:1$, a flavour ratio of $1:1:1$ at Earth, an astrophysical spectrum measured from~\cite{Mohrmann:2015mgs}, and cross-sections according to Eq.~(\ref{eq:gr}) and ~\cite{CooperSarkar:2011pa}. The effects of Doppler broadening on the Glashow resonance (GR)~\cite{loewy2014effect} is also taken into account. \label{fig:rec_nuenergy} }
\end{figure}

Given the negligible atmospheric background rate, the remainder of this Article assumes the event originated from a single high-energy astrophysical neutrino interaction. The major backgrounds to the Glashow resonance are charged-current (CC) interactions (mediated by the exchange of a virtual $W^{\pm}$) of electron (anti)neutrinos with nucleons. Neutral-current (NC) interactions (mediated by the exchange of a virtual $Z^{0}$) from all three flavours are a secondary background. Fig.\,\ref{fig:rec_nuenergy}b illustrates the expected rate from each interaction channel. The posterior distribution of visible energy, reconstructed assuming a cascade hypothesis for different ice models, has a 68\% highest-probability-density region of \depe{} and is shown in Fig.\,\ref{fig:rec_nuenergy}a. Assuming a single power-law astrophysical flux with $\bar{\nu}_{e}:\nu_{e}=1:1$, astrophysical spectral index $\gamma = -2.49$ and normalization at 100\,TeV of \SI{2.33e-18}{GeV^{-1}cm^{-2}s^{-1}sr^{-1}}~\cite{Mohrmann:2015mgs}, we expect to observe 1.55 Glashow resonance hadronic cascades in our data.

Assuming the best-fit flux in ref.~\cite{Mohrmann:2015mgs}, a likelihood-ratio test based on the visible energy rejects both CC and NC interactions in favour of Glashow
resonance with a $P$ value of 0.01, corresponding to a (one-sided) significance of $2.3\sigma$. 
Systematic uncertainties due to the ice modelling and the global energy scale, which affect the visible energy reconstruction, are included. The $P$ value is also tested against spectral assumptions under a single power-law flux, and the results for other spectra are given in Methods. The test’s sensitivity is due to the fact that the visible
energy distribution from Glashow resonance differs both in shape and in normalization from the background at these energies.

This is a conservative estimate which does not rely on early pulses. As muons are produced in meson decay,  the energy of the hadronic shower is directly related to the leading muon energy. In an electron neutrino CC interaction at these energies, on average only about \SI{20}{\percent} of the total neutrino energy is deposited hadronically. 
Thus, while the amount of early Cherenkov light is consistent with the leading muon energy expected from a hadronic shower at the Glashow resonance (\SI{6.3}{\peta \eV}), it is an order of magnitude above that expected from a CC electron neutrino interaction at those energies. In NC interactions, the outgoing neutrino escapes undetected and carries away a large portion of the total energy. Thus, while an NC shower is purely hadronic, a much higher incoming neutrino energy is required. The steeply falling power-law flux of astrophysical neutrinos results in suppression of the NC background.

Although we would ideally incorporate early pulses for CC and NC background rejection, there are several technical challenges that this can pose, including full resimulations of the MC sets that include systematic uncertainties of the hadronic interaction models.
Such studies are under way, and inclusion of this information will be especially important for IceCube-Gen2~\cite{Aartsen:2014njl}, which, owing to its much larger effective area, will record many more events at the Glashow resonance.

\begin{figure}[htp]
\centering
\includegraphics[width=.8\textwidth]{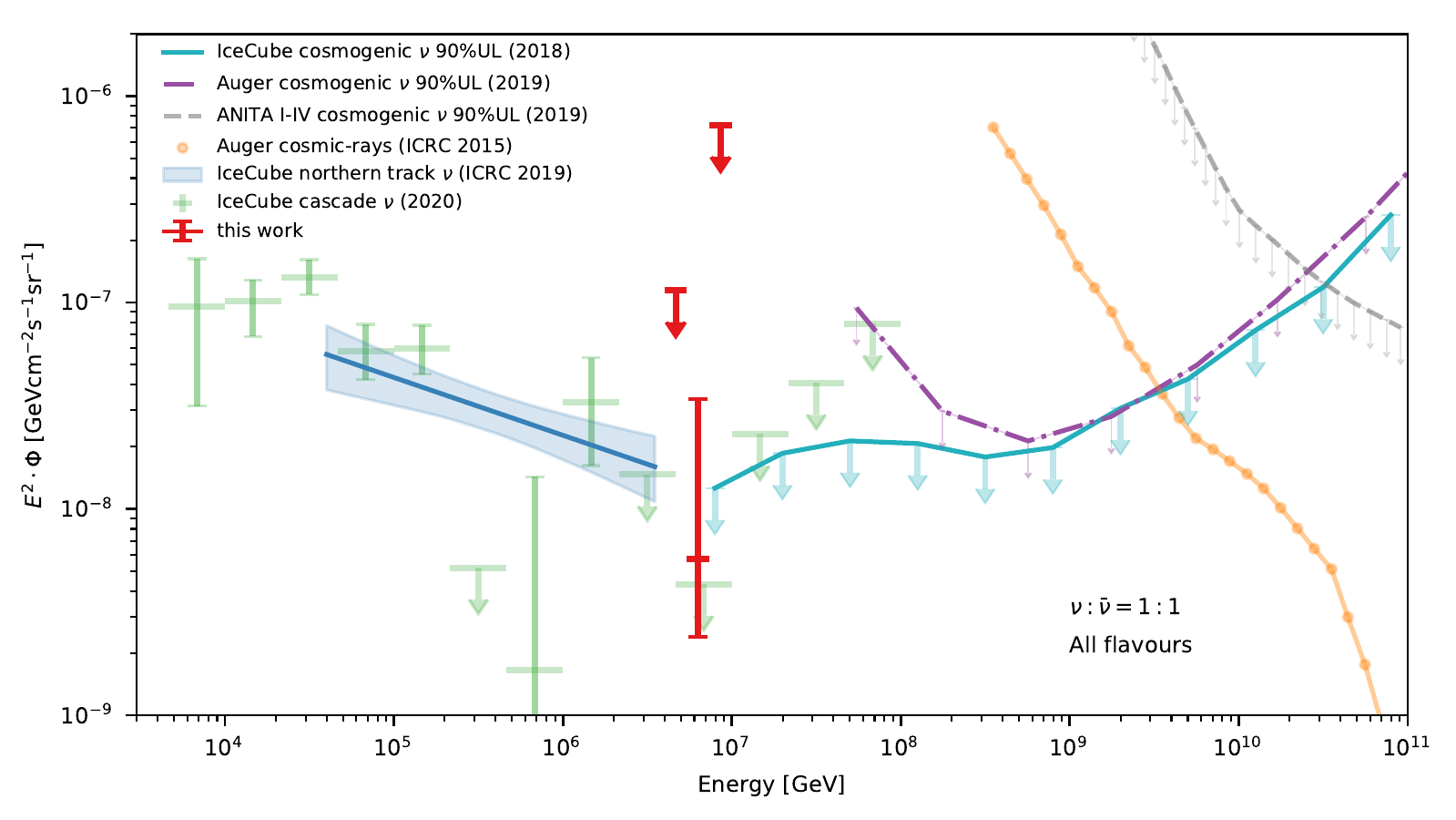}
\caption{\textbf{Measured flux of astrophysical neutrinos.} Global picture of astrophysical neutrino flux measurements \cite{Stettner:2019tok,Aartsen:2020aqd}, cosmogenic neutrino limits \cite{Aartsen:2018vtx,Aab:2019auo,Gorham:2019guw} and ultra-high energy cosmic-ray (UHECR) spectrum \cite{Valino:2015zdi}. 
The $y$ axis is given in term of the energy, $E$, squared times the flux, $\phi$. We assume the ratio $\nu:\bar{\nu} = 1:1$, a flavour ratio of 1:1:1 at Earth, an astrophysical spectrum measured from~\cite{CooperSarkar:2011pa}.
This result extends the measured astrophysical flux to 6.3\,PeV. The luminosity densities of high energy neutrinos and extragalactic ultra-high-energy cosmic rays are found to be comparable.
\label{fig:Global_spectrum}}
\end{figure}

A segmented differential flux fit~\cite{Aartsen:2014gkd} was also performed using three equal-width bins in the logarithm of the neutrino energy over the range
\SI{4}{\peta\eV} to \SI{10}{\peta\eV}. The results, shown in Fig.\,\ref{fig:Global_spectrum} (red), complement other IceCube diffuse analyses~\cite{Stettner:2019tok,IceCube:2020wum,Aartsen:2018vez,Aartsen:2020aqd}. 
The central energy bin extends the measurement of the differential neutrino energy spectrum to 6.3\,PeV, while 68\% upper limits are shown for the lower and upper energy. Arguments based on energetics \cite{Murase:2018utn} and astrophysical unification models such as \cite{Fang:2017zjf,Zhang:2017moz,Muzio:2019leu,Liu:2013wia,Boncioli:2018lrv} suggest a common origin of diffuse gamma-rays, high-energy neutrinos and ultra-high-energy cosmic rays. 
A precise measurement of the cosmic neutrino flux at the Glashow resonance energy will be able to test these predictions, and possibly uncover the origins of ultra-high-energy cosmic rays if the sources can be identified directly via multimessenger observations.

Although the present result focuses on one event, the techniques developed here have implications for the future direction of neutrino astrophysics. 
For example, the idealized $p\gamma$ muon damped model of neutrino production is already inconsistent with the result presented here because such sources produce no electron antineutrinos. With just one event, pp source models cannot be constrained, but the planned IceCube Gen2 experiment~\cite{Aartsen:2014njl} will increase the instrumented volume by an order of magnitude. The statistics collected by such a detector should allow us to differentiate between pp and idealized p$\gamma$ models at a high significance level.

In more realistic source models~\cite{Biehl:2016psj}, multi-pion production in p$\gamma$ sources generate antineutrinos and the $\bar{\nu}_{e}:\nu_{e}$ ratio depends on the photon density, the mass composition of cosmic-rays and also the magnetic field strength of the source. In such cases, a multi-messenger campaign to detect the sources of future Glashow resonance candidates can help determine their production mechanisms. Using the hybrid (early muon and cascade reconstruction) approach could reduce the angular uncertainty by a factor of about $5$, and, as this technique shows, an uncertainty of about $68\,\mathrm{deg}^2$ at $90\%$ containment is possible for hadronic cascades. 
In the near future, such techniques would greatly aid searches for multimessenger counterparts in real time.

\section*{Acknowledgements}
We thank T. Pierog, D. Heck and C. Baus for discussions on realistic hadronic shower simulations in ice. 
We gratefully acknowledge support from the following agencies and institutions: USA—the US National Science Foundation-Office of Polar Programs, the US National Science Foundation Physics Division, the Wisconsin Alumni Research Foundation, the Center for High Throughput Computing (CHTC) at the University of Wisconsin-Madison, the Open Science Grid (OSG), the Extreme Science and Engineering Discovery Environment (XSEDE), the Frontera computing project at the Texas Advanced Computing Center, the US Department of Energy-National Energy Research Scientific Computing Center, the Particle Astrophysics Research Computing Center at the University of Maryland, the Institute for Cyber-Enabled Research at Michigan State University, and the Astroparticle Physics Computational Facility at Marquette University; Belgium—the Funds for Scientific Research (FRS-FNRS and FWO), the FWO Odysseus and Big Science programmes, and the Belgian Federal Science Policy Office (Belspo); Germany—the Bundesministerium f\"ur Bildung und Forschung (BMBF), the Deutsche Forschungsgemeinschaft (DFG), the Helmholtz Alliance for Astroparticle Physics (HAP), the Initiative and Networking Fund of the Helmholtz Association, the Deutsches Elektronen Synchrotron (DESY), and the High Performance Computing Cluster of RWTH Aachen; Sweden—the Swedish Research Council, the Swedish Polar Research Secretariat, the Swedish National Infrastructure for Computing (SNIC), and the Knut and Alice Wallenberg Foundation; Australia—the Australian Research Council; Canada—the Natural Sciences and Engineering Research Council of Canada,
Calcul Qu\'ebec, Compute Ontario, the Canada Foundation for Innovation, WestGrid, and Compute Canada; Denmark—the Villum Fonden, the Danish National Research Foundation (DNRF), the Carlsberg Foundation; New Zealand—the Marsden Fund; Japan—the Japan Society for Promotion of Science (JSPS) and the Institute for Global Prominent Research (IGPR) of Chiba University; Korea—the National Research Foundation of Korea (NRF); Switzerland—the Swiss National Science Foundation (SNSF); the UK—the Department of Physics, University of Oxford.

\section*{Author contributions}
The IceCube neutrino observatory was constructed and is maintained by the IceCube Collaboration. 
A large number of authors contributed to the data processing, detector calibration and MC simulations used in this work. 
The IceCube collaboration acknowledges the substantial contributions to this manuscript from L.L., T.Y. and C. Haack. The final manuscript was reviewed and approved by all authors.

\printbibliography[segment=\therefsegment,check=onlynew]


\newrefsegment

\section*{Methods}
\subsection*{Detection principles}
IceCube instruments a cubic kilometre of Antarctic ice with \num{5160} DOMs, each containing a single downward-facing photomultiplier tube (PMT).
The DOMs are placed on 86 strings that extend from \SI{1450}{\m} to \SI{2450}{\m} beneath the surface~\cite{Aartsen:2016nxy}.
Charged particles travelling above the group velocity of light in ice (velocity $v > 0.73c$) emit light, which is
detectable by the PMTs. 
When multiple groups of PMTs detect photons within a window of \SI{5}{\micro\s}, the recorded charges are digitized and saved as an ``event''~\cite{Aartsen:2016nxy}.

IceCube events are typically classified into two different categories
depending on the outgoing secondaries. 
High-energy muons that start in or travel through the detector are classified as tracks. 
Such muons are created in CC interactions of incoming muon neutrinos. Hadronic and EM showers have shorter extension (about \SI{10}{\m}) and are classified as cascades. 
These can arise either from CC interactions of incoming electron and tau neutrinos, or, with smaller probability, NC interactions of all three neutrino flavours. 
In some cases, tau neutrino CC interactions can produce more complicated event signatures~\cite{Cowen:2007ny}.

Additionally, an interaction via the Glashow resonance can produce
a characteristic event type and energy deposition in IceCube. 
A $W^{-}$ decays to hadrons 67.41\% of the time. 
In such cases, the visible energy is lower than the total energy of all particles in the shower, as there is a substantial fraction of neutral particles and particles with higher energy thresholds to produce Cherenkov radiation~\cite{Aartsen:2013vja}. 
Based on the known properties of the resonance peak—both $M_{W}$ and the relatively narrow width, $\Gamma_{W}$ - the resulting hadronic shower is expected to deposit about 6.0 PeV visible energy. 
Low-energy muons are expected to be produced in and outrun the Cherenkov wavefront of the shower at a rate high enough to potentially trigger early pulses.
In the leptonic channel, the $W^{-}$ decays to an antineutrino and a charged lepton with a branching ratio to each flavour of about 11\%. 
As the antineutrino escapes undetected, a varying fraction of the primary energy is visible as a muon track, EM cascade, or tau. 
This signature is uniquely lacking a hadronic component. 
However, with the wider range of possible charged-lepton energies, it is difficult to distinguish the leptonic channel from CC or NC background without more sophisticated tagging techniques.

\subsection*{Event Selection}

IceCube discovered astrophysical neutrinos using high-energy starting events (HESE) with interaction vertices inside a restricted fiducial
region~\cite{Aartsen:2013bka, Aartsen:2013jdh}. 
These events were selected with a veto-based approach, with the outer layer of the detector used to reject the large muon background. 
The highest-visible-energy cascade detected by HESE was an event of about \SI{2}{\peta\eV} (ref.~\cite{Aartsen:2014gkd}). 
To enhance the detection rate of Glashow resonance events, a larger fiducial volume is required. 
The PEPE selection aims to select multi-PeV, cascade-like events with interaction vertices near the edge of the detector. 
It preferentially selects cascades because of their higher energy resolution compared to that of tracks~\cite{Lu:2017nti}.
At the Glashow resonance, the event rate is about twice that of the HESE selection~\cite{Aartsen:2013jdh}, as shown in Extended Data Table~\ref{table:1}.
The PEPE background is dominated by cosmic-ray muons propagating through corners of the detector that appear cascade-like.

First, events with a total charge of less than \num{1000} photoelectrons or involving fewer than 15 sensors with charge $\geq5$ photoelectrons are rejected to ensure a minimum level of detectable light deposited in the detector to be used in reconstructions. 
Second, events seen previously in the HESE sample~\cite{Aartsen:2013jdh} and tracks in the online extremely high-energy alerts sample~\cite{Aartsen:2018vtx} are removed. 
In addition, events where a single sensor carries more than 50\% of total charge tend to be more difficult to reconstruct and are also excluded. 
Last, to constrain the vertex to be located at the edge of the detector, cumulative charges on the outer two strings are required to be at least  80\% of the total charge. 
After these cuts, the cosmic-ray background rate is about $0.05$ Hz.

To further reject tracks, a BDT was trained against a background of
atmospheric muons, with the signals being $\nu_{e}$ and $\bar{\nu}_{e}$  of neutrino energy \SIrange{1}{10}{\peta\eV}. 
In total, \num{11} features are selected for training as follows: the
charge of the event, the number of triggered sensors above \num{5} photo-electrons, the centre of charge (which describes the location of the event) and its projection onto the x–y plane, the travel of the centre of charge relative to the first quartile in time and its projections onto the z axis and the x–y plane, the ratio of the travel along the z axis to the total number of DOMs, the goodness of fit for a track hypothesis, the time between the first hit overall and the first hit on the sensor with the largest total charge, and the cumulative fraction of total charge at \SI{600}{\nano\s} from the start of the event. 
These parameters summarize the difference between tracks and cascades. 
The charge of the event serves as a proxy for the energy. 
Extended Data Fig.\,\ref{fig:edfig1} shows the BDT score distribution for events with a reconstructed energy above \SI{4}{\peta\eV}. 
The final selection includes only events with a BDT score above \num{0.5} that fit well under a cascade hypothesis~\cite{Aartsen:2013vja}. 
With this, the cosmic-ray background for reconstructed energy $\geq\SI{4}{\peta\eV}$ is about \num{0.02} events per year. 
The BDT score of the event discussed in this work is \num{0.65}.

Nine events passed all selection criteria, with this event the only one having energy above \SI{4}{\peta\eV}. 
The remaining events all had energies below \SI{2}{\peta\eV} and will be discussed in a future publication. 
The effective area, defined as the ratio of the event rate to the flux, near the Glashow resonance energy is increased by a factor of about \num{2} compared to HESE~\cite{Aartsen:2013jdh}.
For \num{4.6} years of data-taking, with the most recent global-fit spectrum~\cite{Mohrmann:2015mgs}, \num{1.55}(\num{0.69}) PEPE events are expected in the hadronic channel of $W$-decay from Glashow resonance assuming $pp$ ($p\gamma$) interactions. 
However, it is worth noting that a spectral cutoff can substantially change the number of expected PEPE events~\cite{Lu:2017nti}.

\begin{etable}
\centering
\begin{tabular}{ |p{2cm}||p{2cm}|p{2cm}|p{2cm}|  }
 \hline
 \multicolumn{4}{|c|}
 {Event Rates per 4.6 yrs} 
 \\
 \hline
$\gamma_{astro}$& -2.28 &-2.49 &-2.89\\
 \hline
 $\Phi_{astro}$ & 4.32 &7.0 &6.45\\
 \hline
 PEPE $pp$   & 2.27	&1.55&	0.28\\
 HESE $pp$ &  1.15&	0.79&	0.14\\
 PEPE $p\gamma$ &1.01&	0.69&	0.12\\
 HESE $p\gamma$  &0.51&	0.35	&0.06\\

 \hline
 \end{tabular}
 \caption{\textbf{Expected event rates for the hadronic decay of $W^{-}$ in IceCube} Expected event rate for the hadronic channel from the Glashow resonance decay using 4.6\,years of data. The spectral indices $\gamma_{\text{astro}}$ are from \cite{Mohrmann:2015mgs,Stettner:2019tok,Schneider:2019ayi} (a recent update~\cite{IceCube:2020wum} has slightly shifted the best-fit values of $\gamma_{\text{astro}}$ and $\Phi_{\text{astro}}$ given in ref.~\cite{Schneider:2019ayi}. 
 The units of $\Phi_{\text{astro}}$ are \SI{e-18}{\per \GeV \per \cm\squared \per \s \per \steradian}.}
    \label{table:1}
\end{etable}

\subsection*{Reconstruction}

The event reported in this Article is reconstructed in a two step process using Bayesian techniques: first to reconstruct the visible energy, interaction vertex, and direction of the initial cascade~\cite{Chirkin:2013}, second using early pulses, with a prior on the interaction vertex~\cite{Haack:2019jpv}.
These two steps rely on independent information, as the early pulses occur on DOMs that are excluded from the initial cascade reconstruction.
Even though the event is not fully contained, its visible energy can still be inferred on the basis of the signature that was detected.

The visible energy is a critical parameter that is highly dependent on the amount of absorption and scattering in the ice surrounding the interaction vertex. 
Cascades are simulated under different hypotheses (vertex position and time, direction and energy), and predicted charge distributions of the DOMs are compared against observed distributions to find those that best match the data.
Systematic uncertainties in bulk scattering and absorption parameters as well as ice anisotropy were taken into account~\cite{Aartsen:2013rt, Chirkin:2019vyq}. 
For the bulk scattering and absorption parameters, the event was reconstructed by scaling the absorption and scattering parameters up and down by \num{5}\%. 
This gave five distinct energies, each corresponding to a different bulk-ice variant. 
A linear regression of these five points gave a model for the reconstructed visible energy as a function of the scattering and absorption parameters. 

Once this was established, an uncorrelated \num{5}\% uncertainty in scattering and absorption was propagated onto the reconstructed visible energy. 
This procedure was repeated for three variations of the ice anisotropy~\cite{Chirkin:2019vyq}, which gave three distinct energy posteriors. These were combined uniformly, assuming equal probabilities for each anisotropy model, to obtain the overall uncertainty on the visible energy due to the ice modelling.
A DOM-to-DOM variation of 10\% is additionally incorporated through a regularization term in the likelihood. 
Finally, a uniform 10\% global energy scale uncertainty is convolved with the previously obtained visible energy posterior probability density function (PDF).

In addition, the  approximate Bayesian calculation (ABC) method~\cite{Chirkin:2013} returns a sample of event vertices and directions for each ice model. 
They are combined across the bulk-ice variants by keeping only samples that have a corresponding test statistic below a pre-computed threshold. Lower values of the test statistic indicate better agreement between data and the resimulated sample. The sample of directions is fitted to a Kent distribution~\cite{kent1982}, while the sample of vertices is passed as a three-dimensional Gaussian prior for the second reconstruction step using early pulses.

The likelihood of observing early pulses at times $t_i^j$ at a given DOM $j$ is given by~\cite{Ahrens:2003fg}:
\begin{align}
    \mathcal{L}(\hat{r},\vec{x}_0,t_0) = \prod_j^{N_\text{DOMs}} \prod_i^{N^j_\text{hits}} \left \{\frac{\alpha_j}{T_j} + (1-\alpha_j)\cdot p(t_i^j \mid \hat{r}, \vec{x}_0, t_0)\right\} \, ,
\end{align}
where $N_{\text{DOMs}}$ is the number of operational DOMs,
$N^j_\text{hits}$ is the number of early pulses at DOM $j$, 
$\alpha_j$ is the expected relative contribution of noise to the hits in DOM $j$,
$T_j$ is the time-window considered for early pulses and $p_j(t_i^j \mid \hat{r},
\vec{x_0}, t_0)$ is the photon arrival time PDF at DOM $j$ for a given track hypothesis with direction $\hat{r}$ and position $\vec{x}_0$ at time $t_0$. 
The photon arrival time PDFs are obtained by interpolating
the results from dedicated MC simulations using spline functions~\cite{Whitehorn:2013nh, Aartsen:2013vja}.
The posterior distribution of the track parameters is given by:
\begin{align}
    p(\hat{r}, \vec{x}_0, t_0 \mid t_i^j) = \frac{\mathcal{L} \cdot p(\hat{r}, \vec{x}_0, t_0)}{\int \mathcal{L} \cdot p(\hat{r}, \vec{x}_0, t_0)~ d\hat{r}d\vec{x}_0dt_0} \, ,
\end{align} 
where the prior $p(\hat{r}, \vec{x}_0, t_0)=p(\hat{r})p(\vec{x}_0)p(t_0)$ assumes independence between groups of parameters. 
The prior on the position $p(\vec{x}_0)$ is taken from vertex posterior of the cascade reconstruction, the prior on the direction $p(\hat{r})$ is a uniform distribution on the sphere, and the prior on the time $p(t_0)$ is a broad Gaussian centered on the vertex time of the cascade reconstruction. 
Samples from the posterior distribution are obtained using nested sampling with the \texttt{dynesty} package~\cite{speagle2019dynesty}, assuming a flat prior for the direction and using the ABC output as a prior for the starting position.
The three blue contours shown in Extended Data Fig.~\ref{fig:Fig2_systematic} correspond to the three anisotropy-model variants, which give different vertex priors.
The muonic component of a hadronic cascade consists of a distribution of muons with different trajectories. The posterior distribution of the track direction is thus to be understood as average direction of the muonic component. Using MC simulations of hadronic cascades, the coverage of the directional posterior was verified with respect to the direction of the initial shower.

\begin{edfigure}[htp]
\centering
\includegraphics[width=.8   \textwidth]{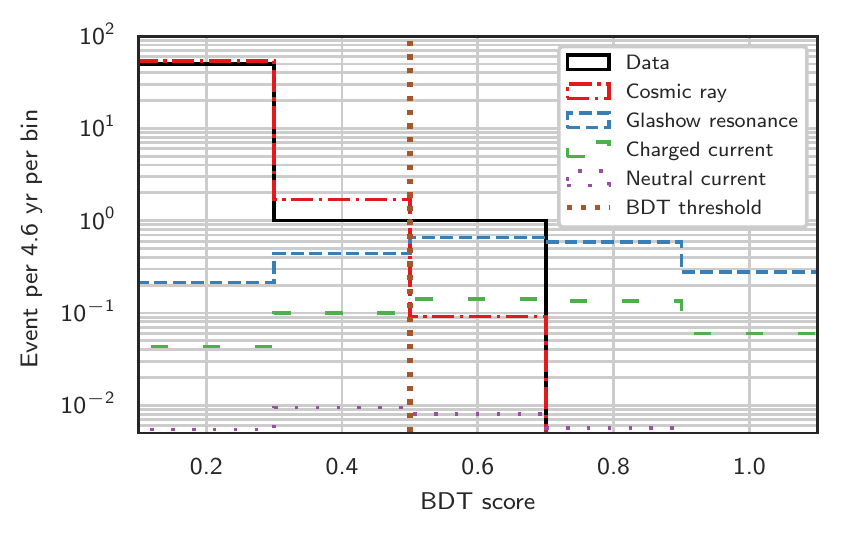}
\caption{
\textbf{The BDT distribution for events with a reconstructed energy above 4 PeV.}
The plotted events are required to be consistent with a
cascade hypothesis based on the goodness of fit. The PEPE event selection
requires a BDT score greater than 0.5. Good data–MC agreements were
observed across the background and the signal region. See Methods for
details. \label{fig:edfig1}
}
\end{edfigure}

\begin{edfigure}[htp]
\centering
\includegraphics[width=.8   \textwidth]{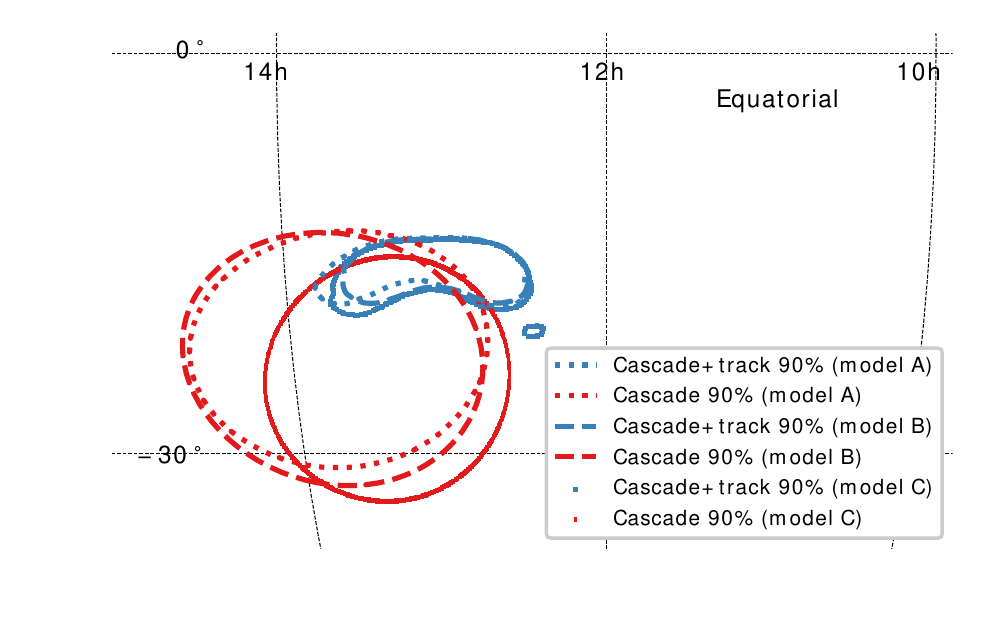}
\caption{
\textbf{The number of strings that observed early pulses for
a given muon energy.}
Shown are reconstructed directions assuming three different ice anisotropy models~\cite{Chirkin:2019vyq} (A, B and C). While the cascade-based reconstructions (red) exhibit some shifts, the hybrid cascade+track reconstructions (blue) appears less susceptible to ice model differences. \label{fig:Fig2_systematic}}
\end{edfigure}

\begin{edfigure}[htp]
\centering
\includegraphics[width=.45 \textwidth, angle=90]{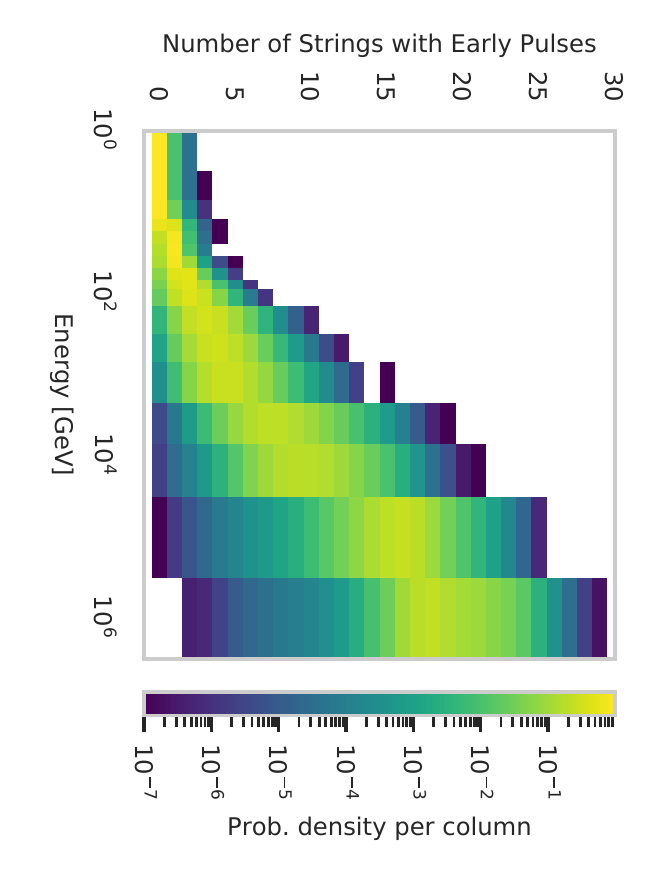}
\caption{
\textbf{The number of strings that observed early pulses for
a given muon energy.}
The colour scale shows the probability of observing the
number of strings with early pulses, $N_{s}$, as function of the simulated muon energy.
\label{fig:p_nstr_vs_e}}
\end{edfigure}

From Extended Data Fig.~\ref{fig:Fig2_systematic}, it is apparent that the cascade reconstruction is more sensitive to ice modelling than the hybrid reconstruction.
Their best fits have a separation of about five degrees. The likelihood
of the hybrid reconstruction is bimodal, with one mode in each lobe.

In order to estimate the energy of the leading muon, we performed MC simulations of \SI{6}{\peta\eV} cascades with a muon of varying energies $E_\mu$. 
The direction and position of the cascade and muon are sampled from the posterior distribution of the early pulse track reconstruction. For every simulated event, the number of strings $N_s$ on which at least one DOM detected early pulses is recorded. This gives the probability of observing $N_s$ strings with early-pulse DOMs for a given $E_\mu$, $\like(N_s | E_\mu)$, shown in Extended Data Fig.~\ref{fig:p_nstr_vs_e}. As $E_\mu$ increases, the muon can penetrate deeper into the detector thus increasing the probability of a larger $N_s$. 

Assuming a uniform prior over $E_\mu$, we can construct $\prob(E_\mu | N_s = 1)$, where $N_s=1$ is what was observed in data, and obtain the \num{68}\% highest posterior density over $E_\mu$. 
Compared to a pure cascade simulation, agreement of simulated first photon arrival times with data is also improved by including such muons, as shown in Extended Data Fig.~\ref{fig:zvt}. 

\begin{edfigure*}[htp]
\begin{subfigure}{0.48\textwidth}
  \centering
  \includegraphics[width=\linewidth]{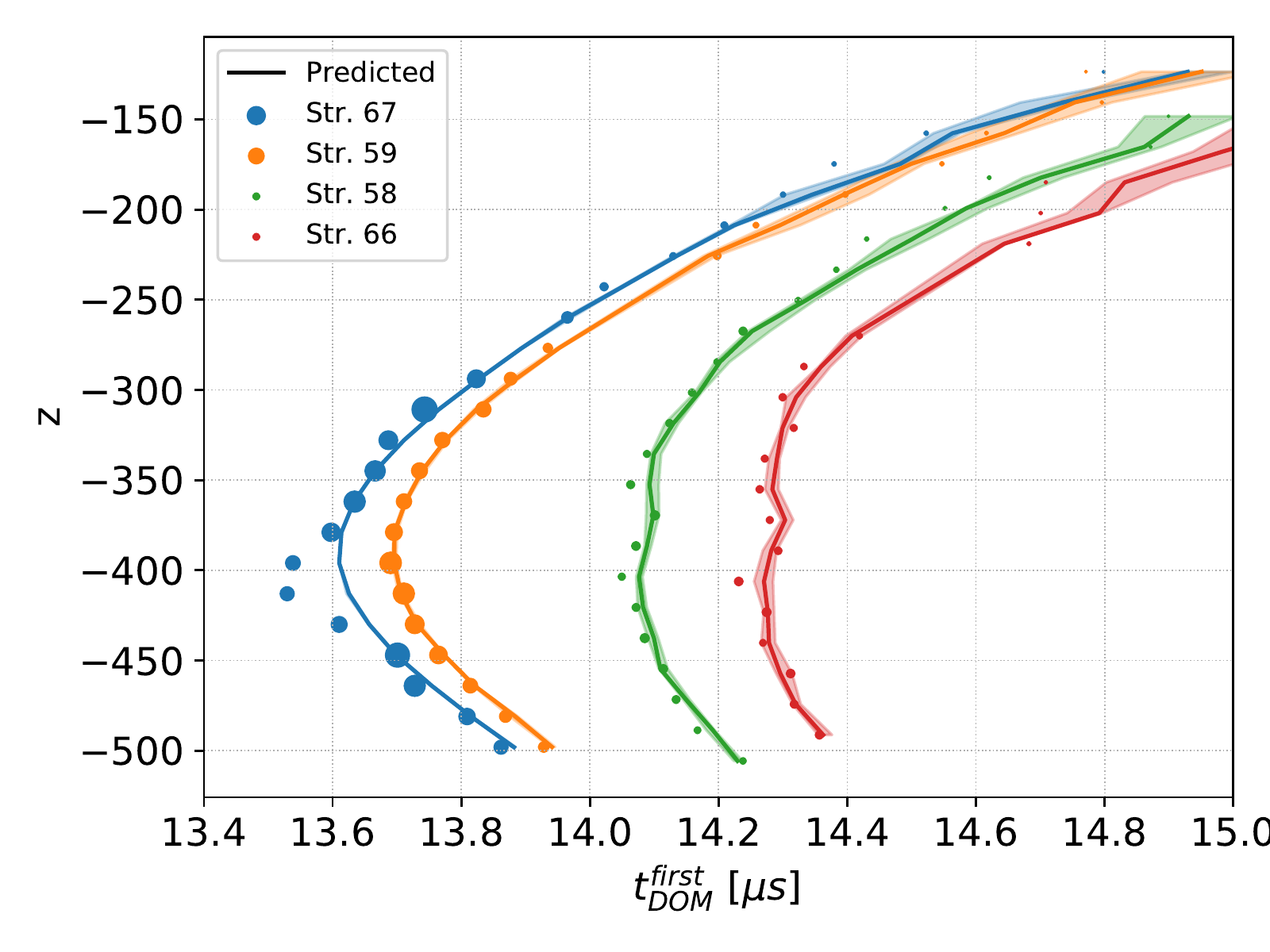}
\end{subfigure}
\begin{subfigure}{0.48\textwidth}
  \centering
  \includegraphics[width=\linewidth]{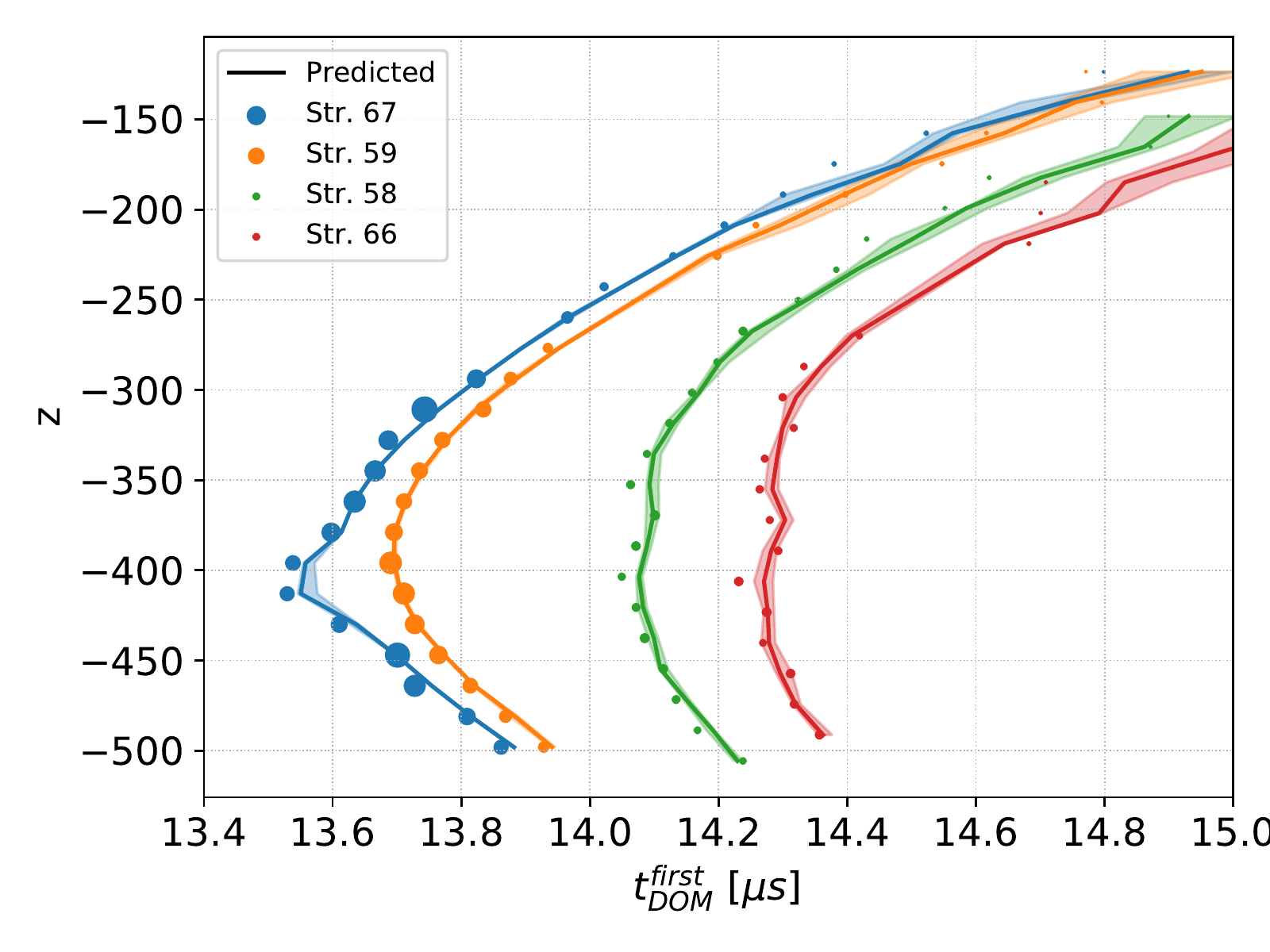}
\end{subfigure}
\protect\caption{
\textbf{First-photon arrival times on four strings.}
Left: First-photon arrival times ($t_{\text{first}}^{\text{DOM}}$) on photosensors deployed on four strings (`Str.', shown in different colours) nearest to the reconstructed vertex plotted against their depth relative to the centre of IceCube (z). 
Observed times are shown as circles, with the size of each circle corresponding to the total charge on that DOM. Predicted arrival times assuming a cascade without escaping muons are shown as lines with shaded regions corresponding to the quartiles obtained from repeated resimulations. Three DOMs on string 67 stand out and have first-photon arrival times that are inconsistent with predictions. 
Right: With the addition of a highly relativistic muon, much better consistency is obtained between observed and predicted first-photon arrival times on the three DOMs with early pulses.}
\label{fig:zvt}
\end{edfigure*}

\subsection*{Atmospheric background rejection}

Combining the two reconstructions above allows for a more significant rejection of the atmospheric background hypothesis than relying solely on the BDT selection. 
For an atmospheric muon to reproduce the event signature seen in the data, it must reach the depth of the detector and suffer a large stochastic loss over a short distance to emulate the initial cascade and have enough energy remaining to deposit early pulses.
We took a numerical approach to calculate a conservative estimate of the number of such events expected over the data-taking period by accepting all such muons that intersect a cylinder of radius \SI{800}{\m} and height \SI{1600}{\m} centred on IceCube. The total muon flux at the surface of the Earth was calculated using MCEq~\cite{Fedynitch:2017M4} with the SIBYLL2.3c hadronic interaction model~\cite{Riehn:2017mfm} and the Gaisser-Hillas 3a primary spectrum~\cite{Gaisser:2013bla}. 
This flux was then propagated to the cylinder using tabulated probabilities-obtained from MC simulations assuming the parameterization for
photonuclear interactions from ref.~\cite{Abramowicz:1997ms} and the parameterization for the bremsstrahlung cross-section from ref.~\cite{Kelner:1995hu} - of a muon of a given energy at the Earth’s surface reaching a particular depth in the ice~\cite{Chirkin:2004hz, Koehne:2013gpa}.
The expectation rate is the surface integral over the vector field of incoming muons. In differential form,
\begin{align}
    dN_\mu &= \like(N_s=1| E_\mu^f) \prob(E_\mu^f | E_\mu^c, \SI{20}{\m}) \like(X | E_\mu^c) \prob(E_\mu^c | E_\mu^i, D(\hat{r},z))\nonumber  \\ 
    &\times \Phi(E_\mu^i, \hat{r}) \hat{r} \cdot d\vec{S} d\Omega dE_\mu^i dE_\mu^c dE_\mu^f, 
\end{align}
where $\Phi(E_\mu^i, \hat{r})$ is the flux of muons with energy $E_\mu^i$ and direction $\hat{r}$ at ground level, $\prob(E_\mu^c | E_\mu^i, D(\hat{r}, z))$ is the PDF for a muon to have energy $E_\mu^c$ after travelling distance $D(\hat{r}, z)$ in ice at to reach depth $z$ given initial energy $E_\mu^i$, $\prob(E_\mu^f | E_\mu^c, \SI{20}{\m})$ the PDF for a muon to have energy $E_\mu^f$ after traveling \SI{20}{\m} (shower length) given initial energy $E_\mu^c$, and $\vec{S}$ denotes the area element of the cylinder. 
The likelihood $\like(X|E_\mu^c)$ is obtained by taking the visible energy posterior and dividing out the most-conservative, uniform prior.

Since the event is downgoing, the expected atmospheric neutrino background is much smaller than the sample average. This is because atmospheric neutrinos at PeV energies are typically accompanied by high-energy muons from the same air shower. These muons should deposit much more energy in the detector than allowed by early pulses. In order to compute the passing fraction, we used Eq.~(3.15) in~\cite{Arguelles:2018awr} with $\prob_\mathrm{light}(E_\mu^f) = 1-\like(N_s=1|E_\mu^f)$. This is then used to down-weight the atmospheric neutrino expectation from the BDT selection.

\subsection*{Calculation of p-value for Glashow resonance}
A likelihood-ratio test under the null hypothesis of exclusively CC or NC astrophysical neutrino interactions was performed. 
As contributions from atmospheric background are negligible, this likelihood-ratio test quantifies the probability that the event did not originate from the Glashow resonance.
 
The likelihood assumes the reconstructed energy is sampled from an inhomogeneous Poisson point process and is given by
\begin{equation}
    \mathcal{L}(S) = e^{-(B+S)}\prod_{i=1}\left \{B\prob_{B}(E_i)+S\prob_{S}(E_i)\right\},
\end{equation}
where $\prob_S(E_i)$ ($\prob_B(E_i)$) is the probability of detecting event $i$ with reconstructed energy $E_i$ under the signal (background) hypothesis, estimated by constructing an unbinned PDF incorporating uncertainties shown in Fig.~\ref{fig:rec_nuenergy}. 
The background normalization, $B$, is the sum of CC and NC expectations in \SI{4.6}{years} of data taking. Assuming $\gamma_{\text{astro}}=2.49$ the estimated background with reconstructed energy between \SI{4}{PeV} and \SI{8}{PeV} is about $0.3$. The signal normalization, $S$, is a free parameter with the null hypothesis being $S=0$. The likelihood-ratio test statistics is then
\begin{equation}
    \Gamma=\log{\frac{\mathcal{L}(S=S_{max})}{\mathcal{L}(S=0)}}.
\end{equation}
Since the number of observed event is one, an analytical solution for maximising the signal probability is
\begin{equation}
    \frac{\partial\mathcal{L}}{\partial S}=0 \Rightarrow {S_{max}=1-B\frac{\prob_{B}}{\prob_{S}}}.
\end{equation}
Pseudo-experiments under the null hypothesis are then generated by randomly sampling a set of pseudo-events from a Poisson distribution, assuming the background normalization, and then sampling the reconstructed energy for each pseudo-event from the background PDF. A $P$ value of $0.01$ was obtained, 
assuming the spectral index $\gamma_{\text{astro}}=2.49$ and $\Phi_{\text{astro}}= 7.0$ given in Extended Table~\ref{table:1},
which corresponds to a rejection of the null hypothesis at (one sided) $2.3\sigma$.

For the softer spectrum with $\gamma_{\text{astro}} = 2.89$, the significance increases to $2.7\sigma$. 
For the harder spectrum with $\gamma_{\text{astro}} = 2.28$, the significance becomes $1.6\sigma$. 
The spectral dependence arises entirely from the deep inelastic scattering background, as the shape of the Glashow resonance is not affected by spectrum assumptions. 
However, a recent measurement of the electron and tau neutrino flux~\cite{Aartsen:2020aqd} has its best-fit spectral index at $2.53$, softer than the global-fit result at $2.49$.

The $P$ value calculation here does not incorporate the early pulse information. To do so requires a fully resimulated and processed sample that includes simulated photons produced by muon secondaries in the shower. 
However, neglecting this information should give a conservative $P$ value estimation as the reconstructed leading muon energy and muon multiplicity are both more consistent with a GR hypothesis than the CC alternative. 
Specifically, assuming the QGSJET-01C hadronic model~\cite{Kalmykov:1997te}, the reconstructed muon energy of \mue{} is more consistent with the quartiles $(20, 37, 72)$\,GeV from a \SI{6}{\peta \eV} hadronic shower than that from a \SI{6.3}{\peta \eV} CC electron neutrino interaction, $(0.04, 1.2, 10)$\,GeV.

\begin{edfigure}[htp]
\centering
\includegraphics[width=.6   \textwidth]{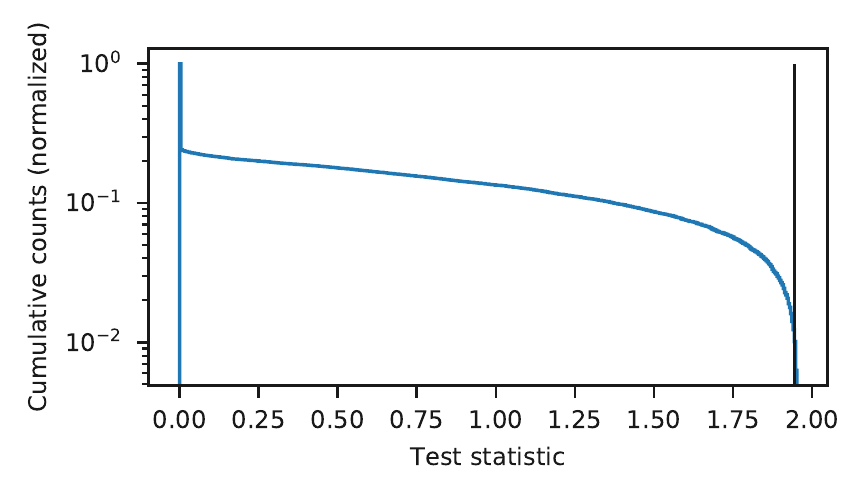}
\caption{
\textbf{The test statistic distribution under the null hypothesis}
Cumulative distribution of $\Gamma$ under the null hypothesis, as generated under the sampling scheme described in the text. The test statistic for the data event is shown in black.}
\label{fig:ex5}
\end{edfigure}

\subsection*{Segmented flux fit} 
The segmented flux fit relies on the MC, which maps true primary energy to reconstructed energy. 
The flux in each of the three true-energy bins is assumed to follow a $E^{-2}$ power law with a floating normalization.
Then, as a function of this flux, the expected event rates in four reconstructed energy bins in the range \SIrange{4}{8}{\peta\eV} were computed assuming Eq.~(\ref{eq:gr}) for the GR cross section and the calculation in~\cite{CooperSarkar:2011pa} for the CC and NC cross sections.
The bin width is chosen to be larger than the energy reconstruction uncertainty. 
The fit is performed over these four reconstructed-energy bins; their expected
event rates are dependent on the flux in the three true-energy bins. 
Comparing the expected event rates to the observation of a single event with reconstructed visible energy \depe{},
we obtain the measured flux normalization and 68\% upper limits shown in Fig.~\ref{fig:Global_spectrum}.
For each true-energy bin, uncertainties were obtained using the Feldman–Cousins construction~\cite{Feldman:1997qc} while treating the normalizations in the other two bins as nuisance parameters that are profiled over.

\subsection*{Data availability}
The full event data, including the location of each DOM, the time and
charge of all pulses associated with this event, and relevant calibration
details are available at \url{https://doi.org/10.21234/gr2021}. 
Additionally, the 90\% contour of the hybrid cascade+track reconstruction shown in Fig.~\ref{fig:rec_direction} and the measured flux shown in Fig. \ref{fig:Global_spectrum} can be found at the same URL.

\subsection*{Code availability}
Much of the analysis code is IceCube proprietary and exists as part of the
IceCube simulation and production framework. IceCube open-source
code can be found at \url{https://github.com/icecube}. 
Additional information is available from \url{analysis@icecube.wisc.edu} upon request.

\printbibliography[segment=\therefsegment,check=onlynew]


\end{document}